\documentclass{article}

\usepackage{PRIMEarxiv}
\usepackage{amsmath}
\usepackage{amssymb}
\usepackage[utf8]{inputenc} % allow utf-8 input
\usepackage[T1]{fontenc}    % use 8-bit T1 fonts
\usepackage[hidelinks]{hyperref}     % hyperlinks
\usepackage{url}      % simple URL typesetting
\usepackage{booktabs}       % professional-quality tables
\usepackage{amsfonts}       % blackboard math symbols
\usepackage{nicefrac}       % compact symbols for 1/2, etc.
\usepackage{microtype}      % microtypography
\usepackage{lipsum}
\usepackage{fancyhdr}       % header
\usepackage{graphicx}       % graphics
\graphicspath{{media/}}     % organize your images and other figures under media/ folder
\title{Financial Frequency Combs}
\author{
  Madhurendra Mishra$^{1}$ \thanks{ORCID: 0009-0003-0080-6186} \\
  Department of Physics \\
  Sri Guru Tegh Bahadur Khalsa College, University of Delhi \\
  New Delhi -- 110007, India \\
  \texttt{madhurendramishra24@gmail.com} \\
  \And
  Armaan Aryan$^{2}$ \\
  Department of Computer Science \\
  Birla Institute of Technology and Science Pilani, Dubai Campus \\
  Dubai International Academic City, Dubai -- 345055, UAE \\
  \texttt{f20220152@dubai.bits-pilani.ac.in} \\
  \And
  Arsh Gogia$^{2}$ \\
  Department of Computer Science \\
  Birla Institute of Technology and Science Pilani, Dubai Campus \\
  Dubai International Academic City, Dubai -- 345055, UAE \\
  \texttt{f20220167@dubai.bits-pilani.ac.in} \\
  \And
  Adarsh Ganesan$^{3}$ \thanks{Corresponding author. ORCID: 0000-0002-5107-8452} \\
  Department of Electrical and Electronics Engineering \\
  Birla Institute of Technology and Science Pilani, Dubai Campus \\
  Dubai International Academic City, Dubai -- 345055, UAE \\
  \texttt{adarsh@dubai.bits-pilani.ac.in} \\
}

\begin{document}
\maketitle

\begin{abstract}
Frequency combs are discrete, equally spaced, phase-coherent spectral lines that emerge from nonlinear mode coupling in physical systems. We show that the incommensurate fractional-order financial model of Huang, Li, Ma, and Chen, whose Caputo derivatives encode macroeconomic long-range memory, generates an analogous structure in its steady-state spectrum. The comb appears only over specific values and ranges of the saving amount $a$, the investment cost $b$, and the demand elasticity $c$, outside which the spectral lines lose their equal spacing. It persists across extended parameter regimes and stays invariant to perturbations in the initial interest rate $x_0$ and investment demand $y_0$, while distinct spectral regimes appear at different initial price levels $z_0$. The comb is generated only when the fractional-order exponents $q_1$, $q_2$, and $q_3$ associated with interest rate, investment demand, and price index are above the critical threshold values. At even higher values of these exponents, the frequency comb transitions into chaos. These findings show that the long-run cyclic structure of a memory-bearing financial economy organises into a discrete, deterministic spectral fingerprint rather than a stochastic continuum.
\end{abstract}

% keywords can be removed
\keywords{Financial Frequency Combs \and Fractional-Order Dynamics \and Caputo Derivative \and Econophysics \and Nonlinear Mode Coupling \and Incommensurate Fractional Orders \and Power Spectrum Analysis \and Bifurcation Analysis \and Macroeconomic Memory \and Chaotic Time Series}

\section{Introduction}
\label{sec:intro}

Over the past three decades, econophysics has developed into a productive bridge between physics and economics~\cite{Mantegna1999,Stanley1996,Chakraborti2011EconophysicsI,Chakraborti2011EconophysicsII,Yakovenko2009Colloquium,Chatterjee2007Kinetic,Smolyak2022ThreeDecades,Kutner2022ThreeRiskyDecades}. By adapting tools originally built for complex physical systems, such as scaling laws, stochastic dynamics, long-range correlations, and nonlinear analysis, researchers have revealed underlying regularities in seemingly erratic market data~\cite{Mantegna1997,Bouchaud2003,Gabaix2003PowerLaw,LuxMarchesi1999,Liu1999Volatility}. Fat-tailed return distributions, multi-fractal scaling, and long-memory correlations are now well documented in stock indices, interest rates, commodity prices, and macroeconomic time series~\cite{Gopikrishnan1999,Plerou1999,Chen1988,Cont2001}. A particularly striking thread is the evidence for a nonlinear deterministic structure in financial time series, though genuine low-dimensional chaos remains contested~\cite{SchienkmanLeBaron1989,Brock1991Nonlinear}. Huang and Li~\cite{HuangLi1993} introduced a canonical model that captures this behavior and later studied in depth by Ma and Chen~\cite{MaChen2001a,MaChen2001b}, who identified coexisting equilibria, periodic orbits, Hopf bifurcations, and chaotic attractors in a three-dimensional system whose variables represent the interest rate, investment demand, and price index. W.C. Chen subsequently extended this model to the fractional-order setting, showing that chaotic dynamics can persist at total derivative orders below~3, even in the presence of memory effects encoded through the fractional derivative order~\cite{Chen2008}.

Frequency combs are discrete, equally spaced, phase-coherent spectral lines that originate in nonlinear optics through mode-locked high-finesse cavities, transforming precision metrology and optical clock technology~\cite{Udem2002,Cundiff2003}. The underlying mechanism, nonlinear mode coupling producing coherent spectral grids, has since proven universal across physical substrates. Phononic frequency combs marked the first extension beyond optics, demonstrated via multimode nonlinear coupling in micromechanical resonators~\cite{Ganesan2017,Ganesan2018a,Ganesan2018b} and subsequently realized across diverse platforms~\cite{Chiout2021,Kesekler2022,anderson2026phononic,deJong2023,Xiao2026}, with theoretical extensions to molecular vibrations~\cite{Lei2024}, twisted van der Waals bilayers~\cite{Liu2025} and engineered solid-state systems~\cite{Rangwala2026}. The comb paradigm extends further to magnonic systems via three-magnon scattering in yttrium iron garnet~\cite{Demidov2020}, terahertz combs from ferroelectric polarization dynamics~\cite{PolarizationComb2026}, and even cosmological settings where exponential quintessence attractors modulate the Hubble parameter~\cite{Trivedi2026} -- pointing to a shared nonlinear dynamical origin across mechanical \cite{Ganesan2017,Ganesan2018a,Ganesan2018b,Chiout2021,Kesekler2022,anderson2026phononic,deJong2023,Xiao2026}, magnonic \cite{Demidov2020}, ferroelectric \cite{PolarizationComb2026}, and cosmological \cite{Trivedi2026} domains.

In this paper, we investigate whether the spectral output of a nonlinear financial model can self-organise into a frequency comb. The fractional-order system of Huang, Li, Ma, and Chen~\cite{HuangLi1993,MaChen2001a,MaChen2001b,Chen2008} governs the co-evolution of interest rate $x$, investment demand $y$, and price index $z$ through a single quadratic nonlinearity. Incommensurate Caputo orders $q_1$, $q_2$, $q_3$ capture the long-range dependence inherent in macroeconomic variables~\cite{West2002,Panas2001}, while structural parameters $a$, $b$, $c$ encode the saving rate, investment cost, and demand elasticity, respectively. We show that the steady-state power spectrum manifests as a financial frequency comb: a set of discrete, nearly-equidistant lines. These combs exist only for specific ranges of structural parameters $a$, $b$, $c$, and fractional-order exponents $q_1$, $q_2$, $q_3$.

\section{Dynamical Model}
\label{sec:model}

\subsection{Integer-Order Financial System}
\label{subsec:integer}
We study the three-dimensional nonlinear financial model of Huang and Li~\cite{HuangLi1993}, later analysed by Ma and Chen~\cite{MaChen2001a,MaChen2001b}, whose state variables are the interest rate $x$, investment demand $y$, and price index $z$:
\begin{align}
  \dot{x} &= z + (y - a)\,x, \label{eq:io1}\\
  \dot{y} &= 1 - b\,y - x^{2}, \label{eq:io2}\\
  \dot{z} &= -x - c\,z. \label{eq:io3}
\end{align}
The parameters $a$, $b$, $c \geq 0$ are the saving amount, investment cost, and demand elasticity, respectively. The interest rate is driven by the investment-savings surplus $(y-a)x$ and the price level $z$; investment demand is sustained by a constant forcing and suppressed by $-x^2$; and the price index relaxes linearly at rate $c$ while being pushed by the interest rate. For $a=3.0$, $b=0.1$, $c=1.0$ the system is dissipative and chaotic, with largest Lyapunov exponent $\lambda_{\max} \approx 0.229$~\cite{Chen2008}.

\subsection{Fractional-Order Generalisation}
\label{subsec:fractional} 

Financial variables carry long-range memory: today's interest rates and prices remain statistically coupled to the distant past~\cite{West2002,Panas2001}. An integer-order derivative sees only the instantaneous rate of change and discards this history entirely. We therefore replace each ordinary derivative in system~\eqref{eq:io1}-\eqref{eq:io3} with a Caputo fractional derivative of order $q_i \in (0,1]$, resulting in a incommensurate system
\begin{align}
  \frac{\mathrm{d}^{q_1} x}{\mathrm{d}t^{q_1}} &= z + (y - a)\,x, \label{eq:fo1}\\
  \frac{\mathrm{d}^{q_2} y}{\mathrm{d}t^{q_2}} &= 1 - b\,y - x^{2}, \label{eq:fo2}\\
  \frac{\mathrm{d}^{q_3} z}{\mathrm{d}t^{q_3}} &= -x - c\,z. \label{eq:fo3}
\end{align}
The Caputo derivative of order $q \in (0,1)$ is defined as~\cite{Caputo1967,Podlubny1999}
\begin{equation}
  \frac{\mathrm{d}^{q} f}{\mathrm{d}t^{q}}
  = \frac{1}{\Gamma(1-q)}\int_0^t (t-\tau)^{-q}\,\dot{f}(\tau)\,\mathrm{d}\tau,
  \label{eq:caputo}
\end{equation}
where $\Gamma(\cdot)$ is the Euler gamma function. The power-law kernel $(t-\tau)^{-q}$ weights past increments of $f$, so the derivative at time $t$ depends on the entire history of the solution. As $q \to 1$ the kernel collapses and the ordinary derivative is recovered. When all three orders are equal ($q_1 = q_2 = q_3 \equiv q$) the system is \emph{commensurate}; the general case is \emph{incommensurate}.

\subsection{Numerical Simulation Setup}
\label{subsec:numerics}

Equations~\eqref{eq:fo1}-\eqref{eq:fo3} are numerically integrated using the Adams-Bashforth-Moulton predictor-corrector scheme proposed by Diethelm~\cite{Diethelm2002}. A fixed step size of $h = 0.01$ is used over the interval $t \in [0, 2000]$. The first 1500 time units are discarded to remove transient behaviour, so the spectral analysis and phase portraits reported below are computed only over $t \in [1500, 2000]$. This window length sets the frequency resolution of the spectra to $0.002$ in inverse time units.

Unless otherwise noted, the parameters are fixed at $a = 3.0$, $b = 0.1$, $c = 1.0$, with initial conditions $x_0 = 2.0$, $y_0 = 3.0$, $z_0 = 2.0$ and fractional orders $q_1 = 1.0$, $q_2 = 1.0$, $q_3 = 0.35$. In each sweep below, one quantity is varied at a time across 1000 evenly spaced values while the rest are held at these baseline settings.

For the fractional-order sweeps, $q_1$ and $q_2$ are each varied between $0.5$ and $1.5$ with the companion order fixed at one and $q_3$ set to $0.3$. When $q_3$ itself is varied, between $0.1$ and $1.0$, $q_1$ and $q_2$ are both fixed at $1.0$. The Fourier spectrum of $X$ is examined in each of these three cases.

The economic parameters $a$, $b$, and $c$ are then swept at the reference fractional orders, over the ranges $[1.0, 5.0]$, $[0.02, 0.2]$, and $[0.5, 1.5]$ respectively. For these sweeps, the spectra of $X$, $Y$, and $Z$ are plotted as frequency-parameter contour maps, which makes it possible to track how the dominant frequency shifts across each range.

Lastly, the initial conditions are perturbed one at a time, with $x_0$ and $y_0$ ranging from $1$ to $10$ and $z_0$ from $1$ to $5$. The Fourier spectrum of $X$ is again used, this time to check that the comb structure persists regardless of the starting point chosen.

\section{Results and Discussion}

\subsection{Financial Frequency Comb Spectrum}

\begin{figure}
    \centering
    \includegraphics[width=0.6\linewidth]{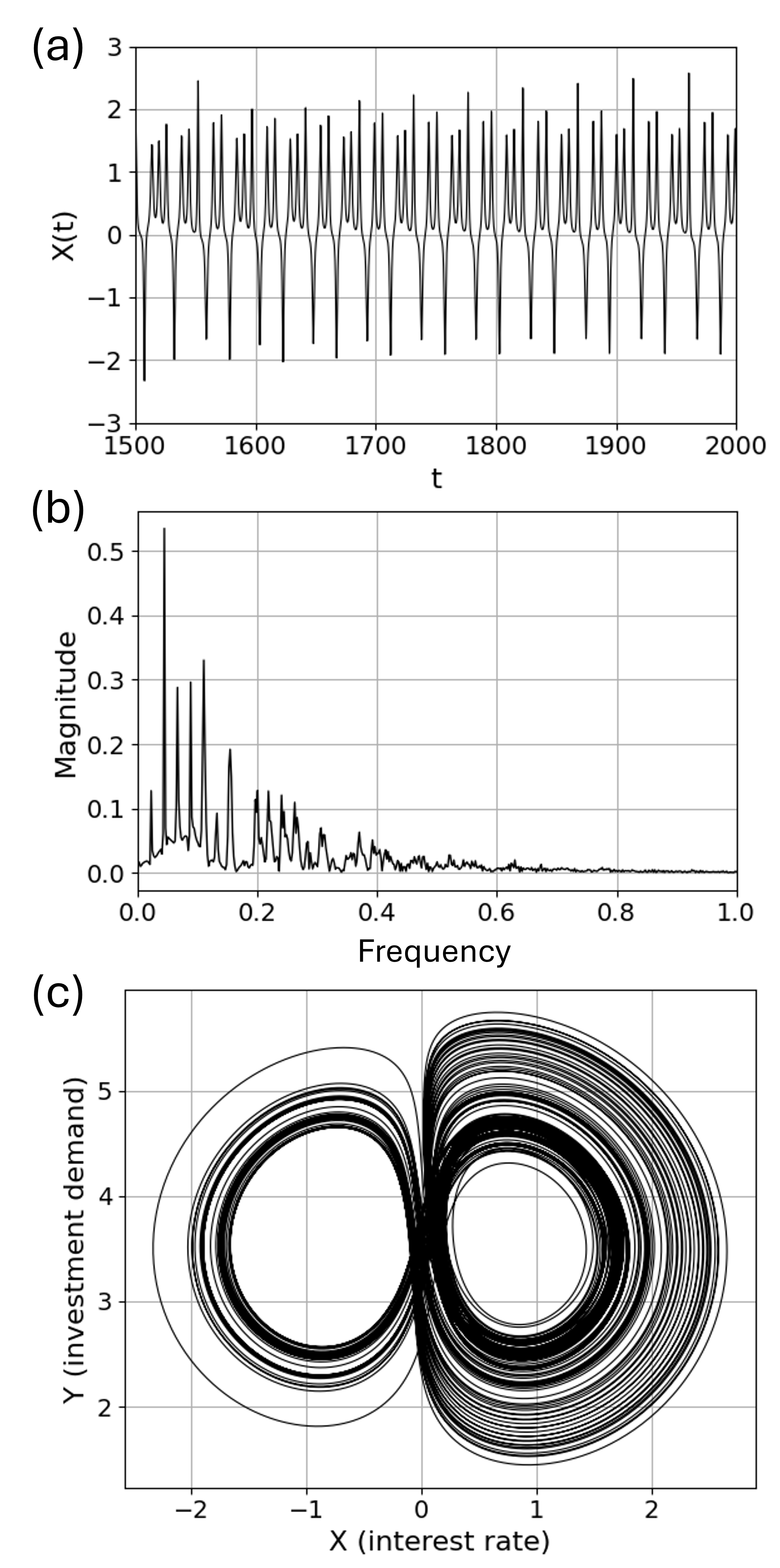}
    \caption{Baseline dynamics at $(q_1,q_2,q_3)=(1.0,1.0,0.35)$,
    $(a,b,c)=(3.0,0.1,1.0)$, $(x_0,y_0,z_0)=(2.0,3.0,2.0)$. (a)~Steady-state time series of $X(t)$. (b)~Normalised FFT magnitude spectrum of $X(t)$. (c)~$X$-$Y$ phase portrait.}
    \label{fig:spiral}
\end{figure}

Fig.~\ref{fig:spiral}(a) shows the long-time behaviour of $X(t)$: this signal shows irregular, burst-like oscillations with structured peaks appearing over time. While the waveform exhibits repeated spike patterns, the spacing, amplitude, and timing are not strictly periodic. Instead, the dynamics appear deterministic but aperiodic, suggesting an underlying nonlinear system operating in a regime between periodic and chaotic behavior, such as intermittent or weakly chaotic oscillations. The FFT spectrum in Fig.~\ref{fig:spiral}(b) shows that the spectral energy does not spread broadly across all frequencies. Below $f\approx0.4$, the spectrum resolves into sharp, uniformly spaced lines with spacing $\Delta f\approx0.05$: the dominant peak lies near $0.05$, with progressively weaker harmonics at $0.10$, $0.15$, $0.20$, and $0.25$. Above this range the signal decays into a weak, featureless background. We define this pattern of discrete, equidistant spectral lines as a \emph{financial frequency comb}. 

The phase portrait (Fig.~\ref{fig:spiral}c) reveals the geometric origin of the comb: the attractor is dense, bounded, and two-lobed. This Lorenz-type trajectory alternates between the two lobes. Each revolution contributes a fast, regular oscillation, while the lobe-switching adds a slower modulation. Both timescales lie well below the Nyquist limit set by the integration step $h=0.01$, so the spectral energy concentrates at the lowest frequencies of the available range. The equidistant comb lines are therefore the harmonic fingerprint of this single bounded two-lobed motion, not the combined signature of several independent oscillators.

\subsection{Economic Parameter Sweeps}

\begin{figure*}
    \centering
    \includegraphics[width=\linewidth]{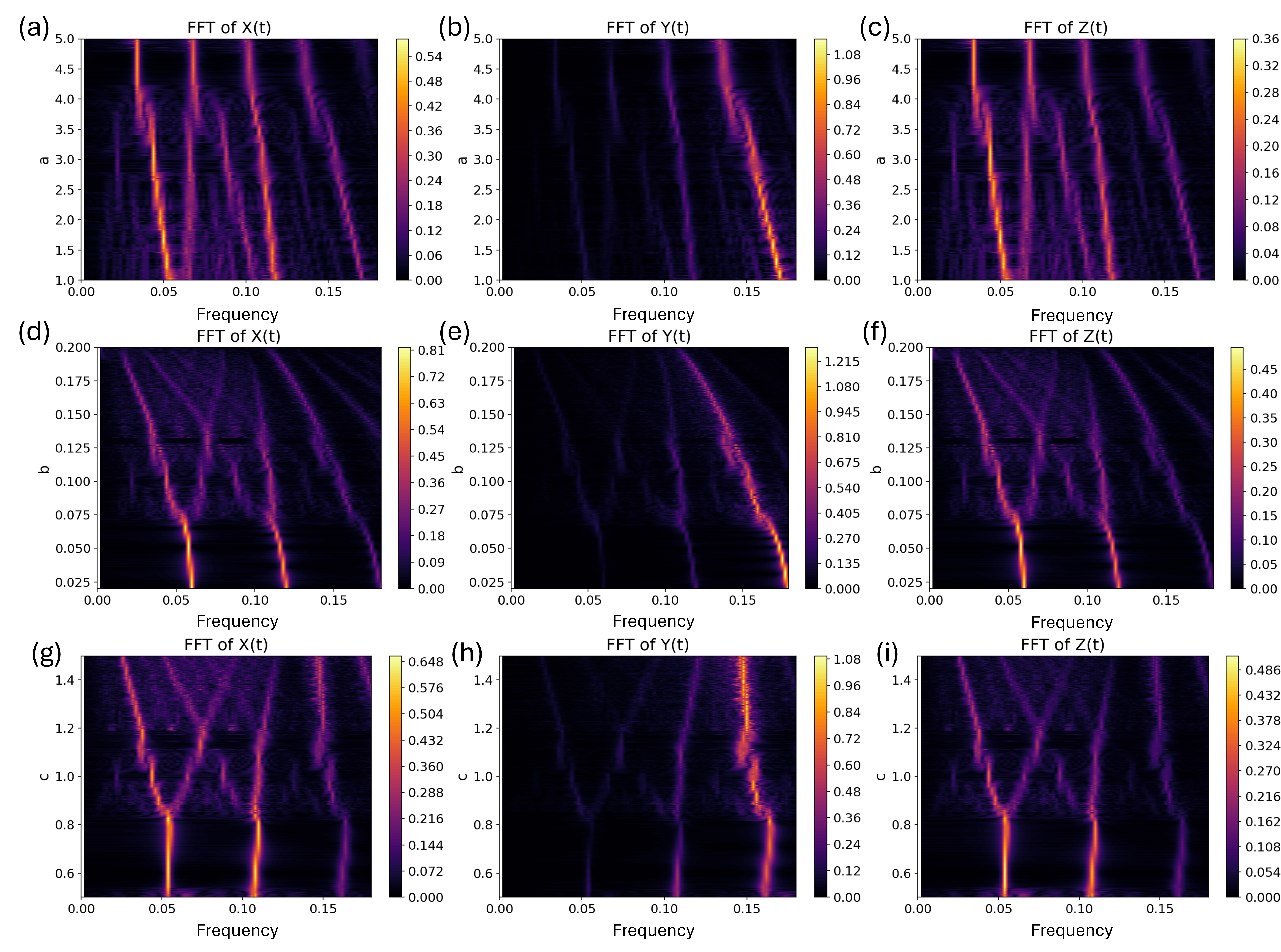}
    \caption{FFT-magnitude spectrograms of $X(t)$ (left column), $Y(t)$ (centre column), and $Z(t)$ (right column), with $(q_1,q_2,q_3)=(1.0,1.0,0.35)$ held fixed. Top row (a)-(c): saving amount $a\in[1.0,5.0]$. Middle row (d)-(f): investment cost $b\in[0.02,0.20]$. Bottom row (g)-(i): demand elasticity $c\in[0.5,1.5]$.}
    \label{fig:comb}
\end{figure*}

Each of the three economic parameters $a$, $b$, and $c$ is varied independently with the fractional orders fixed at $(q_1,q_2,q_3)=(1.0,1.0,0.35)$ (Fig.~\ref{fig:comb}). All three sweeps show the same structural split: $X(t)$ and $Z(t)$ always evolve together spectrally, while $Y(t)$ remains distinct. This follows directly from the model equations: $X$ and $Z$ are coupled through the terms $z+(y-a)x$ and $-x-cz$, so a change in one propagates almost immediately to the other, whereas $Y$ evolves through the term $1-by-x^2$, driving the system via the nonlinear $x^2$ term provide distinct spectral features in comparison to those of $X$ and $Z$.

\subsubsection*{Saving amount $a$}
The spectral structure of $X(t)$ and $Z(t)$ depends strongly on $a$. For $a\gtrsim3.0$ the lines near $f\approx0.05$, $0.10$, $0.15$ are equidistant, constituting a well-defined frequency comb with peak magnitudes near $0.54$ and $0.36$. As $a$ decreases below $3.0$, some of the lines shift to lower frequencies and a few to higher frequencies resulting in unequal frequency spacing. For $a\gtrsim4.25$, the spectral lines do not shift and a stable frequency comb structure is preserved.

\subsubsection*{Investment cost $b$}
For $b\lesssim0.05$, $X(t)$ and $Z(t)$ display equidistant lines, forming a clean frequency comb, with the magnitude of $X$ reaching $\approx0.81$ here, the highest in this sweep. As $b$ increases beyond $\approx0.06$-$0.07$, the lines curve toward each other and eventually cross, forming a chevron pattern in the spectrogram by $b\approx0.10$-$0.13$. When the lines cross, the equal spacing is observed. However, outside of this sweet spot, the lines are no longer equidistant and the frequency combs do not exist; beyond $b\approx0.15$ the lines merge into a broad, irregular band spanning roughly $0.03$-$0.17$ with no discernible periodicity -- hence losing the  phase-coherent harmonic structure.

\subsubsection*{Demand elasticity $c$}
For $c\lesssim0.80$, $X(t)$ and $Z(t)$ exhibit equidistant lines forming a frequency comb. As $c$ increases beyond $\approx0.85$, the lines converge toward each other in a symmetric inverted-V pattern, and the spacing between them decreases. Here again, the spacing is only equal at specific values. When $c\gtrsim1.1$, the lines get merged or broadened into an irregular spectral pattern.

\subsection{Initial-Condition Sensitivity}

\begin{figure}
    \centering
    \includegraphics[width=0.6\linewidth]{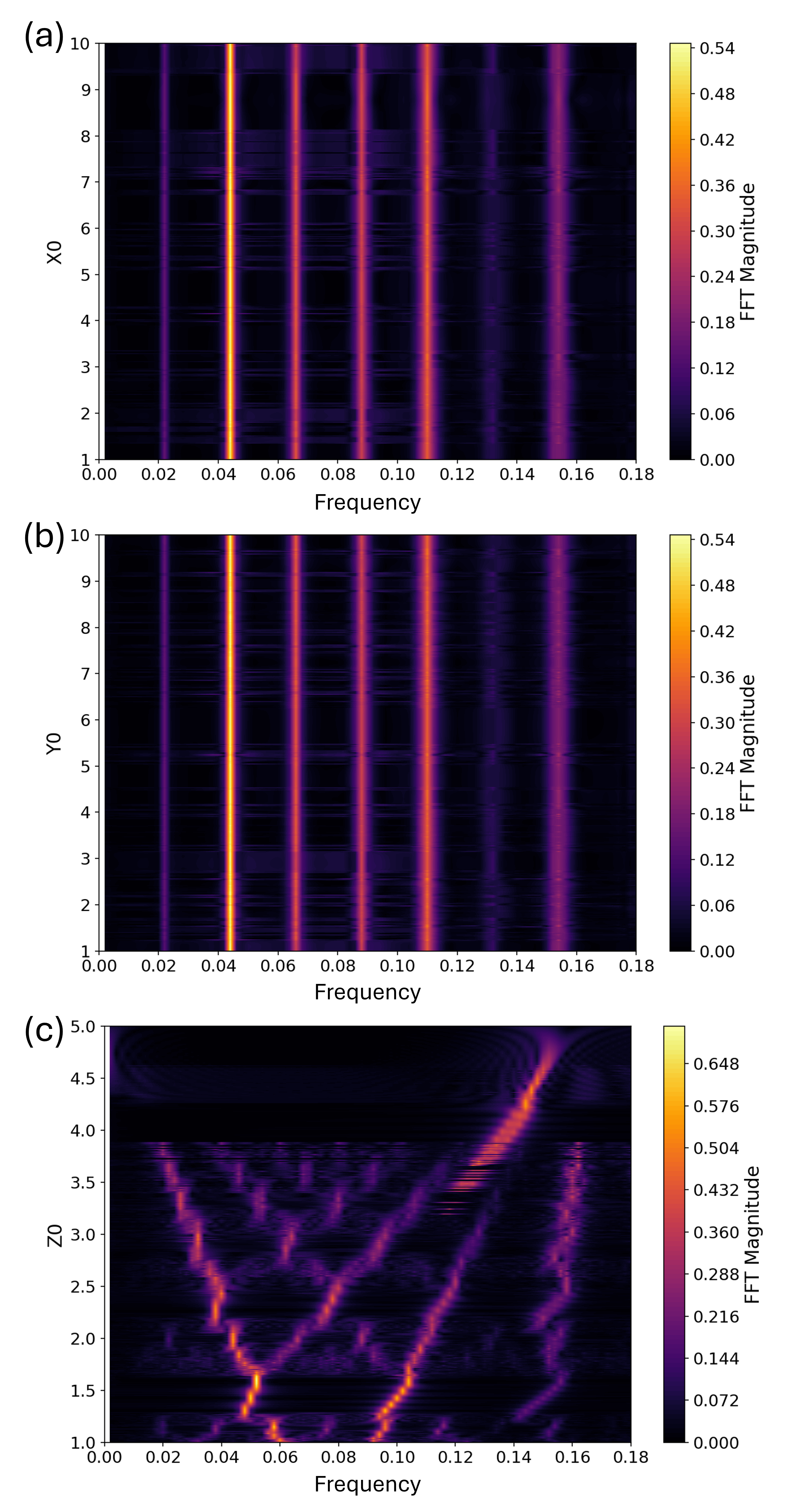}
    \caption{FFT-magnitude spectrograms of $X(t)$ across initial conditions, with $a=3.0$, $b=0.1$, $c=1.0$, and $q_1=q_2=1.0$, $q_3=0.35$ held fixed. Panel~(a): $x_0\in[1,10]$, with $y_0=3.0$ and $z_0=2.0$ fixed. Panel~(b): $y_0\in[1,10]$, with $x_0=2.0$ and $z_0=2.0$ fixed. Panel~(c): $z_0\in[1,5]$, with $x_0=2.0$ and $y_0=3.0$ fixed.}
    \label{fig:IC}
\end{figure}

Next, we investigate the sensitivity of the comb structure to initial conditions by varying $x_0$, $y_0$, and $z_0$ one at a time; Figs.~\ref{fig:IC}(a) and (b) show the sweeps of $x_0$ and $y_0$ over $[1,10]$. In both panels, a comb of seven perfectly vertical equidistant lines appears between $f\approx0.02$ and $0.16$, with uniform spacing $\Delta f\approx0.02$ and dominant magnitude $\approx0.54$; line positions do not shift as $x_0$ or $y_0$ changes. Transients excited by different values of $x_0$ or $y_0$ decay before the analysis window begins, so the long-time spectral structure is entirely set by the attractor geometry.

Fig.~\ref{fig:IC}(c), sweeping $z_0\in[1,5]$, tells a qualitatively different story: for $z_0$ between $1.0$ and approximately $3.8$, the spectrum shows regularly spaced branches spanning $f\approx0.02$-$0.16$. As $z_0$ increases, the discrete jumps in the equidistant spacing of the frequency comb spectrum are observed at discrete values of $z_0$; as $z_0$ increases toward $\approx3.5$, they spread and cross in a complex, non-equidistant pattern, and this is not a frequency comb, since the lines are neither uniformly spaced nor stable in position. At $z_0\approx3.8$-$4.0$ the frequency comb spacing increases five-fold with the fundamental tone appearing at $f\approx0.13$ for $z_0=4.0$. As $z_0$ increases further, the fundamental frequency increases linearly with its value reaching $0.16$ at $z_0=5.0$ (magnitude $\approx0.65$).

\subsection{Fractional-Order Sweeps}

\begin{figure}
    \centering
    \includegraphics[width=0.6\linewidth]{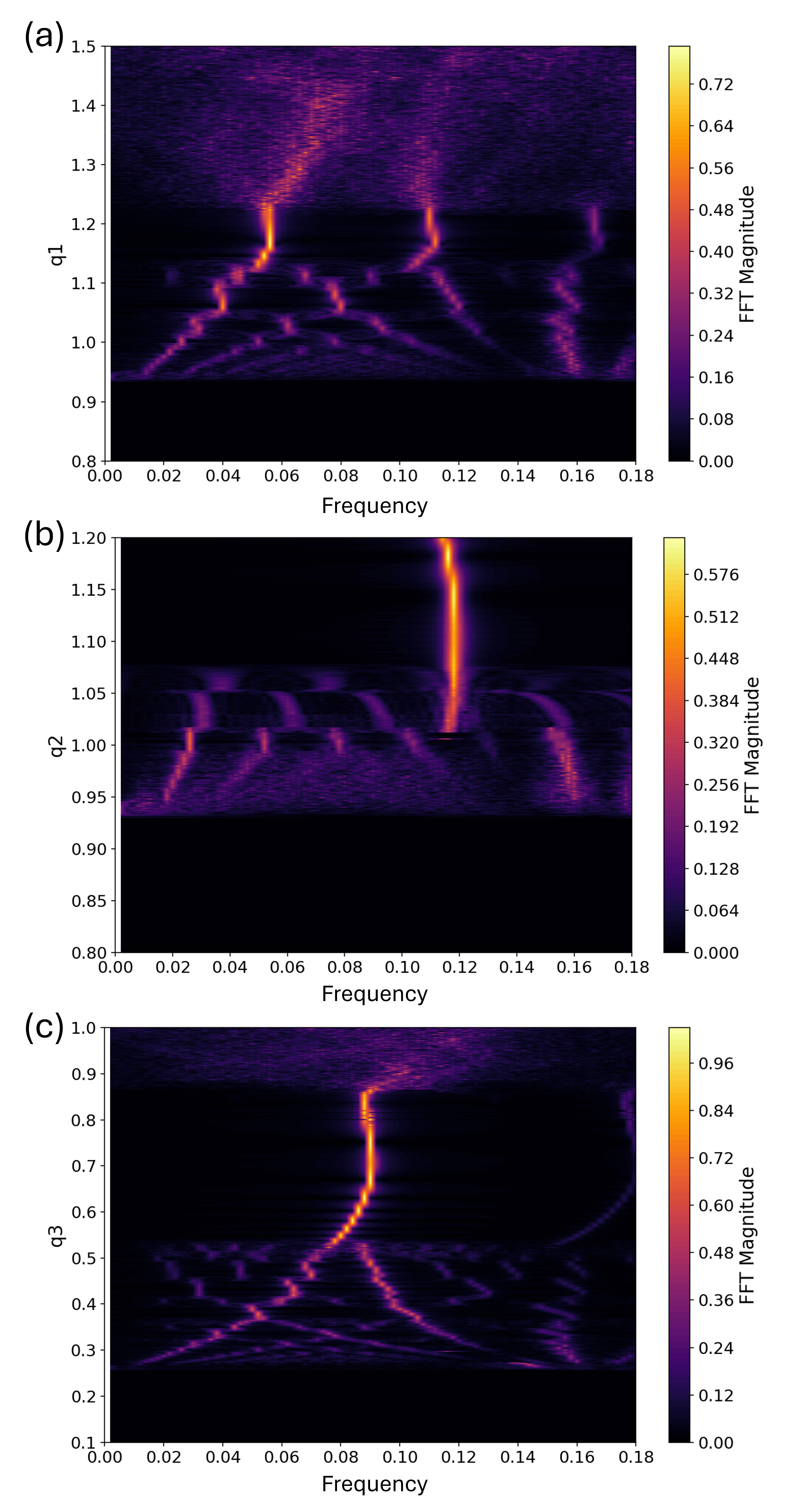}
    \caption{FFT-magnitude spectrograms of $X(t)$ across fractional orders, with $a=3.0$, $b=0.1$, and $c=1.0$ held fixed throughout. Panel~(a): $q_1\in[0.5,1.5]$ (displayed from $0.8$ to $1.5$), with $q_2=1.0$ and $q_3=0.3$ fixed. Panel~(b): $q_2\in[0.5,1.5]$ (displayed from $0.80$ to $1.20$), with $q_1=1.0$ and $q_3=0.3$ fixed. Panel~(c): $q_3\in[0.1,1.0]$, with $q_1=q_2=1.0$ fixed.}
    \label{fig:qsweep}
\end{figure}

Sweeping $q_1$, $q_2$, and $q_3$ independently with $(a,b,c)=(3.0,0.1,1.0)$ fixed (Fig.~\ref{fig:qsweep}) gives a clear picture of how fractional memory controls the spectral structure.

\subsubsection*{Interest-rate order $q_1$}
Below $q_1\approx0.95$, there is no self-sustained oscillation: the spectral magnitude is negligible throughout the frequency range. Above this threshold, the system begins to oscillate and multiple spectral lines appear. These lines are equidistant throughout the oscillatory range $q_1\approx0.95$-$1.25$, so the pattern qualifies as a frequency comb over this entire window. However, the equidistant spacing is not constant: it increases progressively with $q_1$, so the comb extends out to higher frequencies as interest-rate exponent grows. Discrete jumps in line position are also observed at several values of $q_1$, reflecting abrupt transitions between distinct resonant states of the attractor. Within the narrower window $q_1\approx1.15$-$1.22$ the comb lines are sharpest and best-defined, with peaks near $f\approx0.055$ and $0.110$ and peak magnitude $\approx0.72$. Above $q_1\approx1.25$ the dynamics enter the chaotic regime: the lines broaden, merge, and lose their equidistance, and the spectral content becomes irregular. Interest-rate exponent therefore does not improve spectral coherence monotonically: the comb exists over a well-defined range of $q_1$, is most organised within a narrower sub-window, and degrades into chaos above $q_1\approx1.25$.

\subsubsection*{Investment-demand order $q_2$}
Below $q_2\approx0.93$ there is no self-sustained oscillation. Between $q_2\approx0.93$ and $1.00$ multiple spectral lines appear at low frequencies ($f\lesssim0.10$) leading to a frequency comb. At $q_2\approx1.00$ a discrete transition occurs: the frequency comb transitions abruptly to a single strong spectral ridge near $f\approx0.12$, which persists with stable magnitude $\approx0.58$ across $q_2\in[1.00,1.18]$ without developing a multi-line pattern. While a single spectral line is not essentially a comb, the presence of its weak harmonics also justifies for its classification as a 
'frequency comb'.

\subsubsection*{Price-index order $q_3$}
The $q_3$ sweep traces the spectral evolution of $X(t)$ across three distinct regimes. Below $q_3\approx0.27$ there is no self-sustained oscillation. Between $q_3\approx0.27$ and $0.50$, the spectrum is filled with regularly spaced branches spanning $f\approx0.02$-$0.16$. By $q_3\approx0.50$-$0.55$, this structure condenses into a single narrow ridge near $f\approx0.085$-$0.09$. Together with its weak harmonics, it can be called a 'frequency comb'. As $q_3$ increases further, the ridge intensifies and bends slightly toward higher frequency, reaching peak magnitude ($\approx0.96$) across $q_3\approx0.65$-$0.80$. Above $q_3\approx0.82$, the ridge broadens, shifts to near $f\approx0.095$, and enters the chaotic phase. The price-index fractional order $q_3$ thus governs the full spectral evolution: insufficient memory prevents any oscillation from arising; moderate memory produces a frequency comb; and excess memory drives the system into irregular, multi-line chaotic dynamics with no equidistant spacing.

\section{Conclusion}
\label{sec:conclusion}

We have shown that the Huang-Li-Ma-Chen financial system of fractional-order, under suitable parameter and memory-order conditions, settles in a discrete frequency comb generated by the single nonlinear coupling between the interest rate $x$ and the investment demand $y$. The comb is unaffected by the initial interest rate $x_0$ and investment demand $y_0$, yet reorganises completely when the initial price index $z_0$ is changed, a direct consequence of the long memory carried by this variable and a dynamical analogue of price stickiness in real economies. Its existence is bounded rather than generic, appearing only within finite ranges of the saving rate $a$, the investment cost $b$, the demand elasticity $c$, and the three fractional-order exponents $q_1$, $q_2$, $q_3$; outside these ranges, the oscillation either decays or its harmonics broaden into a continuous, incommensurate spectrum. Between these limits lies a narrow regime of organized chaotic dynamics, where the motion remains aperiodic overall, yet retains enough order around a near-periodic core to preserve an evenly spaced harmonic ladder.

The main finding of this work is that long-memory financial systems may possess deterministic spectral fingerprints, called 'frequency combs'. If similar signatures are found in real macroeconomic or financial time series, they could become powerful tools for detecting hidden economic cycles, estimating memory effects, forecasting regime changes, validating nonlinear economic models and supporting policy analysis. Since the comb exists only for specific ranges of saving rate, investment cost, demand elasticity and fractional memory orders. the observed spectral combs in economic data could help estimate these underlying structural parameters. Since there exists a progression from no oscillation, irregular broadband spectrum, regular frequency comb and chaotic broadband spectrum, monitoring spectral evolution could therefore provide an early warning indicator of transitions between stable and unstable economic regimes. Instead of validating financial models only using time-series statistics, researchers can compare frequency spacing, harmonic amplitudes, comb stability, between simulations and observed macroeconomic data to assess whether a model captures realistic long-memory dynamics. Because the comb changes systematically with economic parameters, policymakers could use it to study how changes in saving policies, investment incentives, price-control mechanisms, interest-rate regulations affect long-term cyclical behaviour of an economy.

%Bibliography
\bibliographystyle{unsrt}  
\bibliography{references}

\end{document}